\documentclass[a4paper,11pt]{article}
\usepackage{pos}
\usepackage{float}
\usepackage{booktabs}
\usepackage{makecell}
\usepackage{caption}
\usepackage{physics}
\newcommand{\Op}{\mathcal{O}}

\title{\begin{flushright}
\small{
WUB/25-02\\
CERN-TH-2025-026}
\vskip 0.7cm
\end{flushright}
Flavor mixing in charmonium and light mesons with optimal distillation profiles}
\ShortTitle{Flavor mixing in charmonium and light mesons with optimal distillation profiles}

\author*[a]{Juan Andrés Urrea-Niño}
\author[b]{Jacob Finkenrath}
\author[a]{Roman Höllwieser}
\author[a]{Francesco Knechtli}
\author[a]{Tomasz Korzec}
\author[c]{Michael Peardon}

\affiliation[a]{Department of Physics, Bergische Universität Wuppertal,\\
Gaußstraße 20, 42119 Wuppertal, Germany}

\affiliation[b]{CERN,\\
Esplanade des Particules 1, 1211 Geneva 23, Switzerland}

\affiliation[c]{School of Mathematics, Trinity College Dublin,\\
Dublin 2, Ireland}

\emailAdd{urreanino@uni-wuppertal.de}
\emailAdd{hoellwieser@uni-wuppertal.de}
\emailAdd{korzec@uni-wuppertal.de}
\emailAdd{knechtli@uni-wuppertal.de}
\emailAdd{j.finkenrath@cern.ch}
\emailAdd{mjp@maths.tcd.ie}

\abstract{We study the light meson - charmonium - glueball mixing using flavor-singlet meson operators built from optimal distillation profiles together with purely gluonic operators in different $J^{PC}$ channels at two different pion masses ($\approx$ $420$, $800$ MeV) in two $N_{\rm f} = 3 + 1$ ensembles at close to physical charm quark mass. We observe non-zero mixing correlations between the different types of operators and quantify the overlaps between states created by them and the energy eigenstates by means of a GEVP formulation. We are particularly interested in the scalar glueball and its possible decay into two pions so we also include two-pion operators in our calculation.}

\FullConference{The 41st International Symposium on Lattice Field Theory (LATTICE2024)\\
 28 July - 3 August 2024\\
Liverpool, UK\\}

\begin{document}
\maketitle

\section{Introduction}
The goal of this work is to continue the study of the flavor- and glueball-mixing in the mapping of the scalar flavor-singlet light meson and charmonium spectrum in $N_f = 3 + 1$ gauge ensembles \cite{Roman-Ensembles,Höllwieser:2019Dd} first presented in \cite{UrreaNino:2024BY}. This includes the application of the improved distillation technique \cite{Urrea, Peardon} to construct operators leading to states with large overlap onto the energy eigenstates of interest. A state of particular interest in this symmetry channel is the scalar glueball, which can decay into even numbers of pions among other possible multi-particle final states. To approach its study systematically we work at two different pion masses $\left( 420, 800\ \text{MeV}  \right)$. Based on the quenched glueball mass estimate of around $1800$ MeV \cite{PeardonGlueballs}, the lighter pion setup allows for the decays into 2 and 4 pions while for the heavier pion only the 2-pion decay is possible. To account for such decays, we also consider 2-pion operators in the basis of operators to map out the scalar spectrum. Non-zero correlations between 1-particle and 2-particle operators, as well as between meson and glueball ones, can provide us information about the composition of the different energy eigenstates and the calculation of the resulting finite-volume energy spectrum is a first step towards a scattering study via the Lüscher method. 

\section{Methods}
Details of the ensemble corresponding to the $800$ MeV pion mass, denoted as A1h, can be found in \cite{UrreaNino:2024BY}. The ensemble corresponding to the $420$ MeV pion mass is denoted by A1 and details can be found in \cite{Roman-Ensembles}. It has lattice size $96\times 32^3$, $\beta = 3.24$, lattice spacing $a = 0.0536 (11)$ fm and includes three degenerate clover-improved Wilson light quarks at the $SU(3)$ flavor symmetric point as well as a close-to-physical charm quark and Lüscher-Weisz gauge action. Both ensembles have open boundary conditions in time. In both ensembles we calculate on every time-slice sufficiently away from the boundaries $N_v = 200$ eigenvectors of the 3D gauge-covariant Laplacian built from 3D APE-smeared gauge links \cite{APE} using the Thick-Restart Lanczos algorithm also used in \cite{Urrea}. The perambulators are built using the solvers available in the open-source package openQCD version 1.6 \cite{openQCD}. The charm perambulators in both ensembles include the 200 Laplacian eigenvectors while for the light perambulators we use 200 eigenvectors for A1h and 100 for A1. This reduction in the number of vectors used is not a major concern since improved distillation has been shown to use the available vectors in the best way possible \cite{ThesisUrrea}. In this work we measure all entries of a mixing correlation matrix for the channel $J^{PC} = 0^{++}$ of the form
\begin{align}
C(t) &= \begin{pmatrix}
C_{cc}(t)  & C_{cl}(t) & C_{cg}(t) & C_{c2\pi}(t)\\
C_{lc}(t) & C_{ll}(t)  & C_{lg}(t) & C_{l2\pi}(t)\\
C_{gc}(t) & C_{gl}(t) & C_{gg}(t)  & C_{g2\pi}(t)\\
C_{2\pi c}(t) & C_{2\pi l}(t) & C_{2\pi g}(t) & C_{2\pi 2\pi}(t)
\end{pmatrix}.
\label{eqn:CorrMatrix}
\end{align}   
Here $C_{q_1 q_2}(t)$ is the correlation between flavor-singlet meson operators $ \left \langle \bar{q}_1(t) \Gamma q_1(t) \cdot \bar{q}_2(0) \tilde{\Gamma} q_2(0) \right \rangle$, where $\tilde{\Gamma} = \gamma_0 \Gamma^{\dagger} \gamma_0$ and $q_1, q_2 \in \{ c,l \}$. The off-diagonal correlations $C_{cl}(t)$ and $C_{lc}(t)$ include the explicit flavor-mixing between light mesons and charmonium. It is labeled "explicit" since flavor-mixing also happens through the disconnected correlations in $C_{cc}(t)$ and $C_{ll}(t)$ independently of these off-diagonal terms, in which case it is called "implicit" mixing \cite{MixingBali}. $C_{gg}(t)$ is the correlation between gluonic operators, which here correspond to the sum of the Laplacian eigenvalues \cite{Morningstar2013}. $C_{2\pi 2\pi}(t)$ is the correlation between the 2-pion operators in the scalar flavor-singlet channel. All remaining off-diagonal entries correspond to the mixing between the different types of operators. In \cite{UrreaNino:2024BY} we presented the details on how to build all entries except for the 2-pion operator. Here we focus on the inclusion of this new type of operator. In light flavor $SU(3)$ the pions are members of the flavor-octet and a product of two such states lives in the product representation which is reducible as $8 \otimes 8 = 1 \oplus 8 \oplus 8^{\prime} \oplus \overline{10} \oplus 10 \oplus 27$. For details on this decomposition, along with the Clebsch-Gordan coefficients required to build a basis for each of the irreducible representations, see \cite{CGCoeffs}. The Wick contractions necessary to calculate $C_{2\pi 2\pi}(t)$ are efficiently calculated with a FORM code \cite{FORM}. Mixing terms involving the 2-pion operator are also calculated in this way. The different topologies of diagrams relevant for the mixing matrix are displayed in Table \ref{table:Diagrams}. All diagrams in a given cell must be linearly combined with pre-factors defined by the Wick contractions. As for flavor-singlet 1-particle correlation functions, the flavor-singlet 2-pion correlation function involves disconnected terms which dominate the overall statistical error. For our first study of the mixing involving the 2-pion operator we consider the case of zero spatial momentum for both pions and use standard distillation, i.e the meson profile is $1$. The energy spectrum is extracted via the GEVP formulation \cite{Luscher, Blossier}. To make this calculation more numerically stable we perform two operations before solving the GEVP. First, we calculate the singular vectors corresponding to the largest singular values of the charmonium and light mesons diagonal blocks, denoted as $V_{c}$ with $3$ columns and $V_{l}$ with $5$ columns. We include more vectors for the light meson operators because we have more states to resolve. We then define a "partially" pruned \cite{Balog, Niedermayer} correlation matrix as
\begin{align}
\tilde{C}(t) &=
\begin{pmatrix}
V_c^{\dagger} & 0 & 0 & 0\\
0 & V_l^{\dagger} & 0 & 0\\
0 & 0 & 1 & 0\\
0 & 0 & 0 & 1 
\end{pmatrix}
\begin{pmatrix}
C_{cc}(t)  & C_{cl}(t) & C_{cg}(t) & C_{c2\pi}(t)\\
C_{lc}(t) & C_{ll}(t)  & C_{lg}(t) & C_{l2\pi}(t)\\
C_{gc}(t) & C_{gl}(t) & C_{gg}(t)  & C_{g2\pi}(t)\\
C_{2\pi c}(t) & C_{2\pi l}(t) & C_{2\pi g}(t) & C_{2\pi 2\pi}(t)
\end{pmatrix}
\begin{pmatrix}
V_c & 0 & 0 & 0\\
0 & V_l & 0 & 0\\
0 & 0 & 1 & 0\\
0 & 0 & 0 & 1 
\end{pmatrix}
\end{align} 
Second, the entries of this new matrix are normalized as $\tilde{C}_{ij}(t) \rightarrow \frac{\tilde{C}_{ij}(t)}{\sqrt{\tilde{C}_{ii}(a) \tilde{C}_{jj}(a)}}$. All the statistical analysis in this work is done using the \textit{pyerrors} library \cite{Joswig2023} which uses the $\Gamma$-method \cite{Wolff2004,Schaefer2011} with automatic differentiation \cite{Ramos2019}.
\begin{table}
\begin{center}
\begin{tabular}{c| c| c| c| c}
\toprule
                  & $\Op_{l}$ & $\Op_{c}$ & $\Op_{2\pi}$ & $\Op_{g}$ \\
\midrule
$\Op_{l}$       &\makecell{\includegraphics[width=1.5cm]{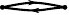}, \\ \includegraphics[width=1.5cm]{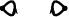}}& \includegraphics[width=1.5cm]{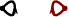}&\makecell{\scalebox{-1}[1]{\includegraphics[width=1.5cm]{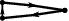}},\\ \scalebox{-1}[1]{\includegraphics[width=1.5cm]{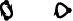}}}&\includegraphics[width=1.5cm]{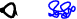}\\
\midrule
$\Op_{c}$ & - &\makecell{\includegraphics[width=1.5cm]{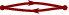},\\ \includegraphics[width=1.5cm]{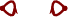}} &\scalebox{-1}[1]{\includegraphics[width=1.5cm]{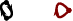}}&\includegraphics[width=1.5cm]{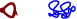}\\  
\midrule
$\Op_{2\pi}$ & - & - & \makecell{\includegraphics[width=1.5cm]{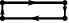}, \includegraphics[width=1.5cm]{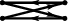},\\ \includegraphics[width=1.5cm]{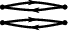},\includegraphics[width=1.5cm]{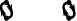}} & \includegraphics[width=1.5cm]{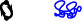}\\
\midrule
$\Op_{g}$         & - & - & - &\includegraphics[width=1.5cm]{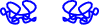}\\
\bottomrule
\end{tabular}
\end{center}
\caption{Diagrams required to calculate the entries of the mixing correlation matrix. Black (Red) lines denote propagation of light (charm) quarks and blue knots denote the glueball operator at a given time.}
\label{table:Diagrams}
\end{table}

\section{Results}
\begin{figure}
    \centering
    \begin{minipage}{0.47\textwidth}
        \centering
        \includegraphics[width=0.95\textwidth]{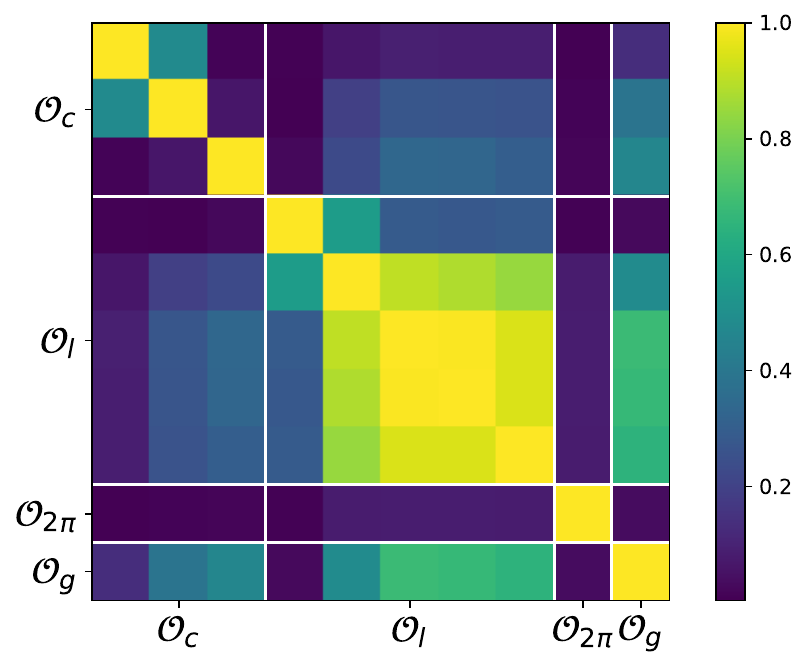} % first figure itself
        \caption{Normalized mixing correlation matrix at time $t=a$ in the ensemble A1.}
        \label{fig:CorrelationsA1}
    \end{minipage}\hfill
    \begin{minipage}{0.47\textwidth}
        \centering
        \includegraphics[width=0.95\textwidth]{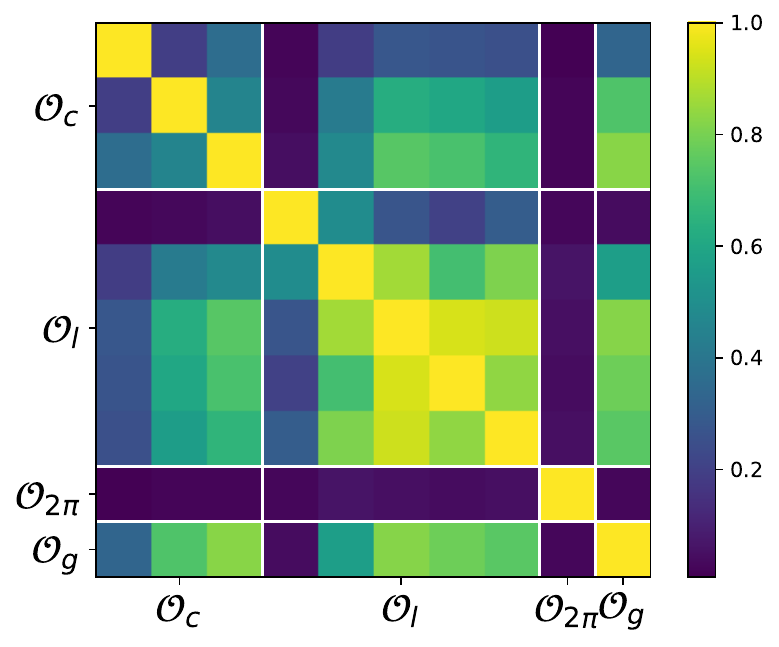} % second figure itself
        \caption{Normalized mixing correlation matrix at time $t=a$ in the ensemble A1h.}
        \label{fig:CorrelationsA1heavy}
    \end{minipage}
\end{figure}

\begin{figure}%[H]
\centering
\includegraphics[width=0.75\textwidth]{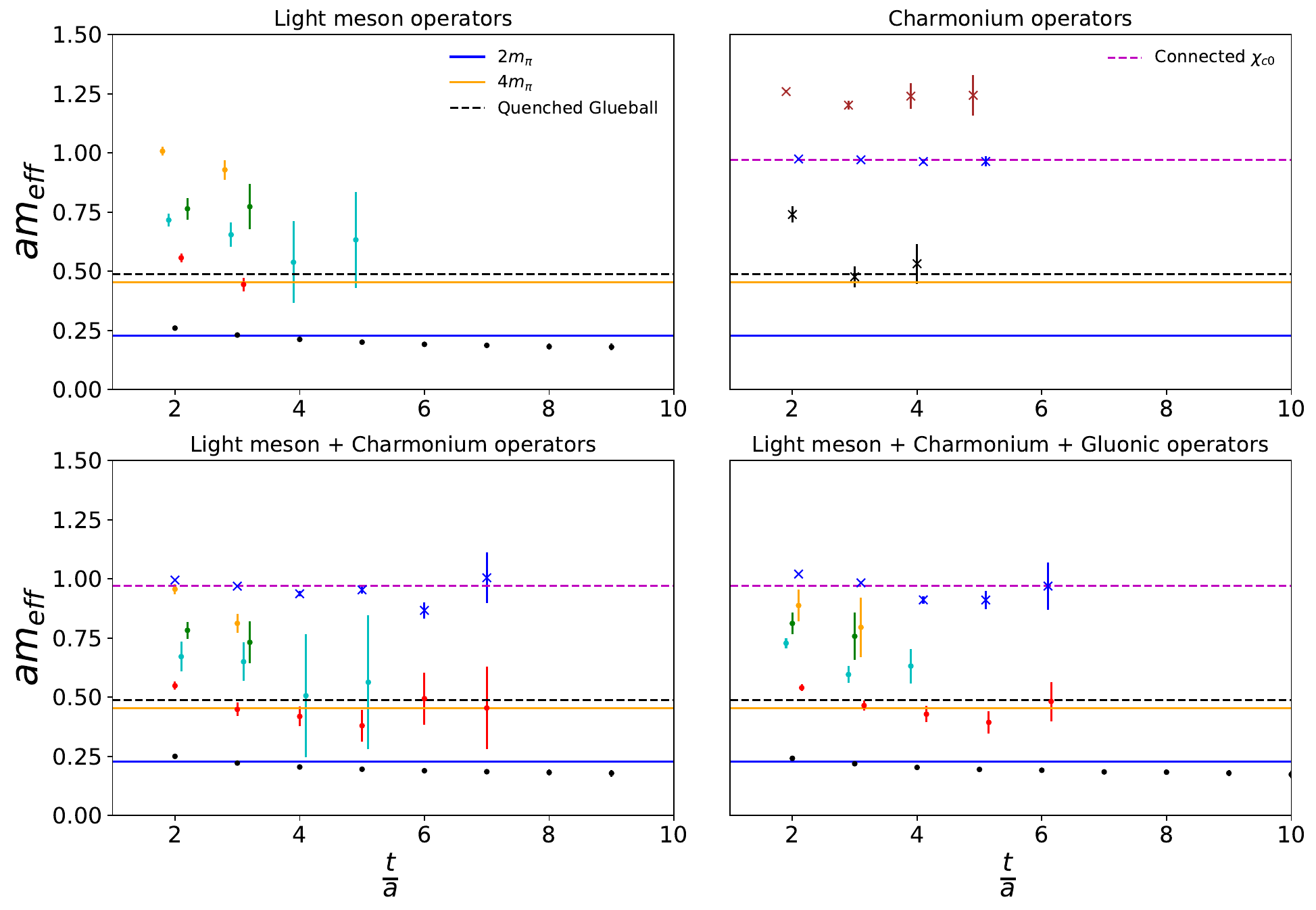}
\caption{Effective masses for $J^{PC} = 0^{++}$ calculated from the different GEVPs involving the mixing between different types of 1-particle operators in ensemble A1. Horizontal lines serve as reference for 2 and 4 times the pion mass together with the quenched scalar glueball prediction. The horizontal magenta line corresponds to the plateau average of the ground-state scalar charmonium calculated neglecting the disconnected contribution to the correlation.}
\label{fig:SpectrumA1_1}
\end{figure}

\begin{figure}%[H]
\centering
\includegraphics[width=0.75\textwidth]{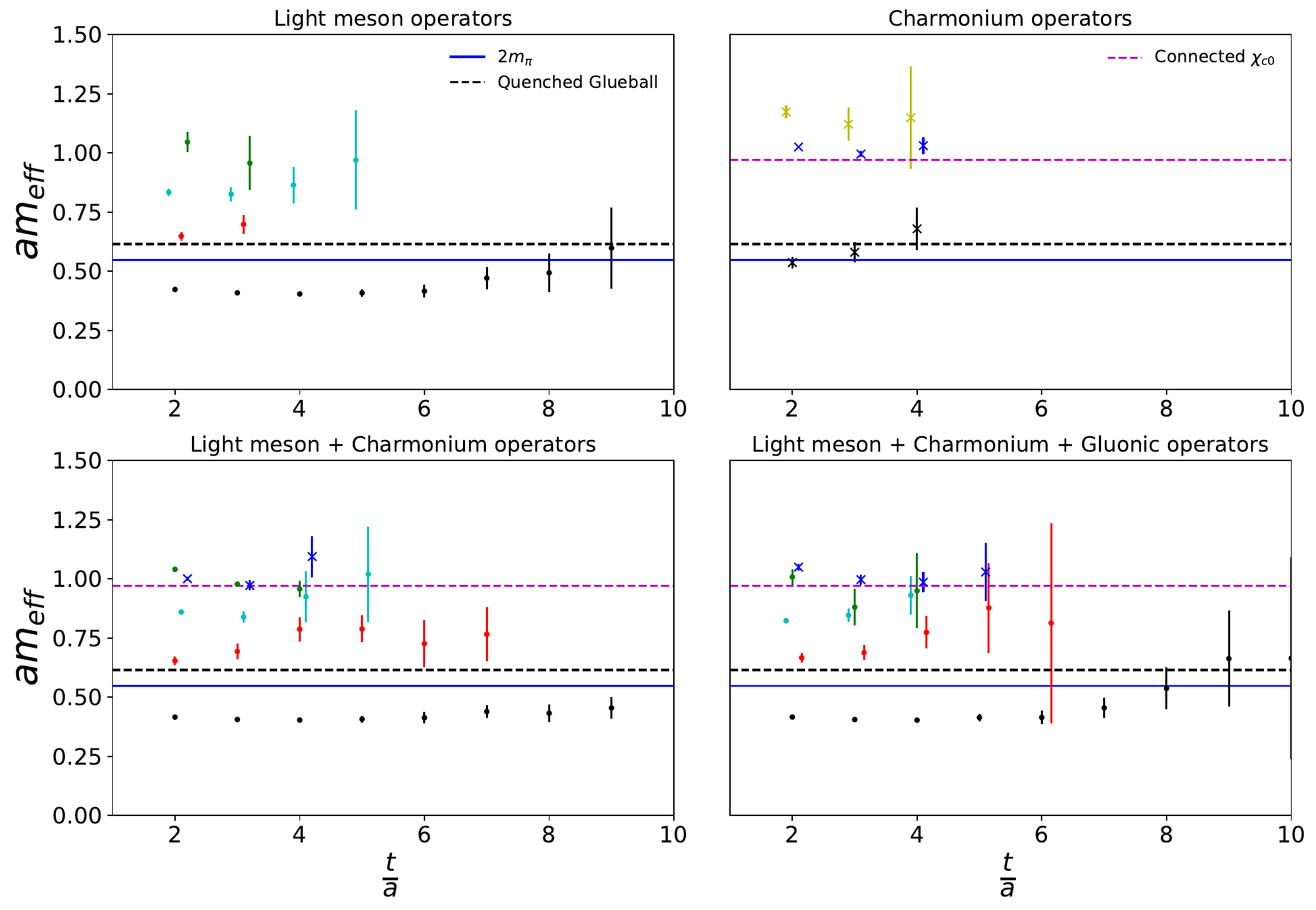}
\caption{Effective masses for $J^{PC}=0^{++}$ calculated from the different GEVPs involving the mixing between different types of 1-particle operators in ensemble A1h. Horizontal lines serve as reference for 2 times the pion mass together with the quenched scalar glueball prediction. The horizontal magenta line corresponds to the plateau average of the ground-state scalar charmonium calculated neglecting the disconnected contribution to the correlation.}
\label{fig:SpectrumA1heavy_1}
\end{figure}

Our first goal is to investigate the off-diagonal entries which correspond to correlations between different types of operators and check if they are non-zero with statistical significance. Figs. \ref{fig:CorrelationsA1} and \ref{fig:CorrelationsA1heavy} show the entries of the correlation matrix at $t=a$ for both ensembles, where due to the chosen normalization all diagonal entries are equal to $1$. Non-zero off-diagonal correlations are observed between the different types of operators, in particular between light mesons and charmonium as well as mesons and the gluonic operator. The former indicates that the explicit flavor-mixing is non-negligible, and therefore worthwhile including, while the latter indicates how purely gluonic and mesonic operators create states which overlap onto common energy eigenstates. This is also indicated in other studies of the scalar and pseudo-scalar channels  \cite{Zhang,Jiang,Frere, Cheng, Cheng2, Janowski}. To study the different types of mixing present in our matrix we calculate the energy spectrum from different sub-blocks of the mixing matrix involving different types of operators. Figs. \ref{fig:SpectrumA1_1} and \ref{fig:SpectrumA1heavy_1} show the calculated effective masses from the GEVP in four different cases for both ensembles: light mesons operators only, charmonium operators only, light meson and charmonium operators together and finally the full correlation matrix. The first feature of interest is the ground state effective masses obtained from the charmonium operators which is in the region of the light meson states, significantly below the mass coming only from the charmonium connected correlation but above the light scalar ground state. Since no explicit flavor-mixing is included in this GEVP, the presence of this light state is exclusively due to the charm-disconnected contribution to the correlations. This clearly indicates states created from charmonium operators have non-negligible overlaps onto light meson states. The first excitation in the spectrum is consistent with the mass extracted from connected correlations only, which hints at it being the true ground charmonium state. This is confirmed when both light meson and charmonium operators are combined together in a larger GEVP: the light state seen by the charmonium operators is absorbed by the light meson ones, most likely the first excited state, while the state suspected to be the ground charmonium state remains consistent with the connected-only result.\\

\begin{figure}%[H]
\centering
\includegraphics[width=0.75\textwidth]{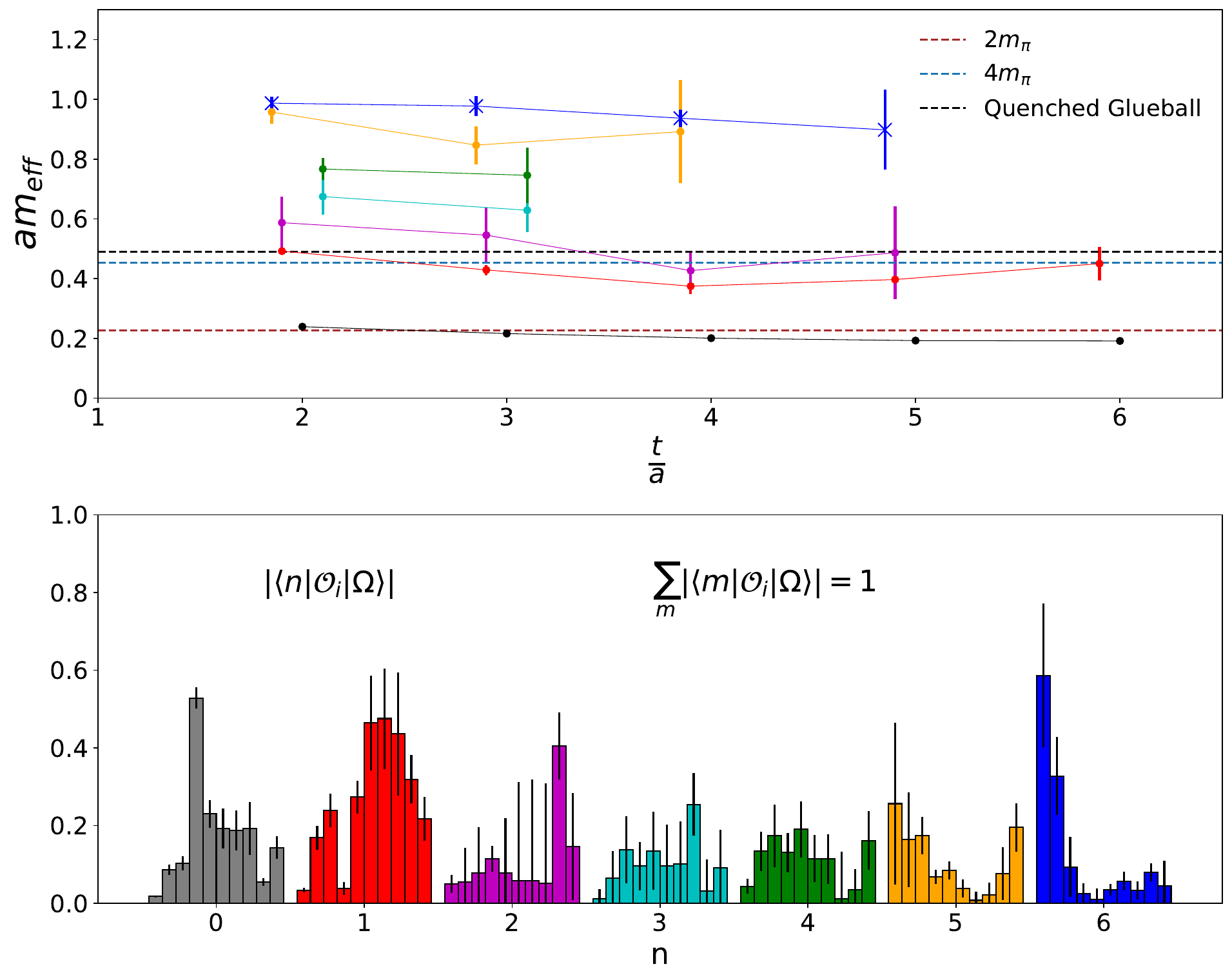}
\caption{\textbf{Upper panel:} Effective masses for $J^{PC} = 0^{++}$  from the GEVP involving all operators from the mixing matrix in ensemble A1. We only show points with reasonable error bars. The connecting lines are drawn to guide the eye. \textbf{Lower panel:} Normalized overlaps between the states created by each operator and each energy eigenstate. Bars of the same color correspond to  overlaps with the same energy eigenstate. The bars correspond in order to 3 charmonium operators, 5 light meson operators, one 2-pion operator and one gluonic operator.}
\label{fig:Spectrum2piA1}
\end{figure}

\begin{figure}%[H]
\centering
\includegraphics[width=0.75\textwidth]{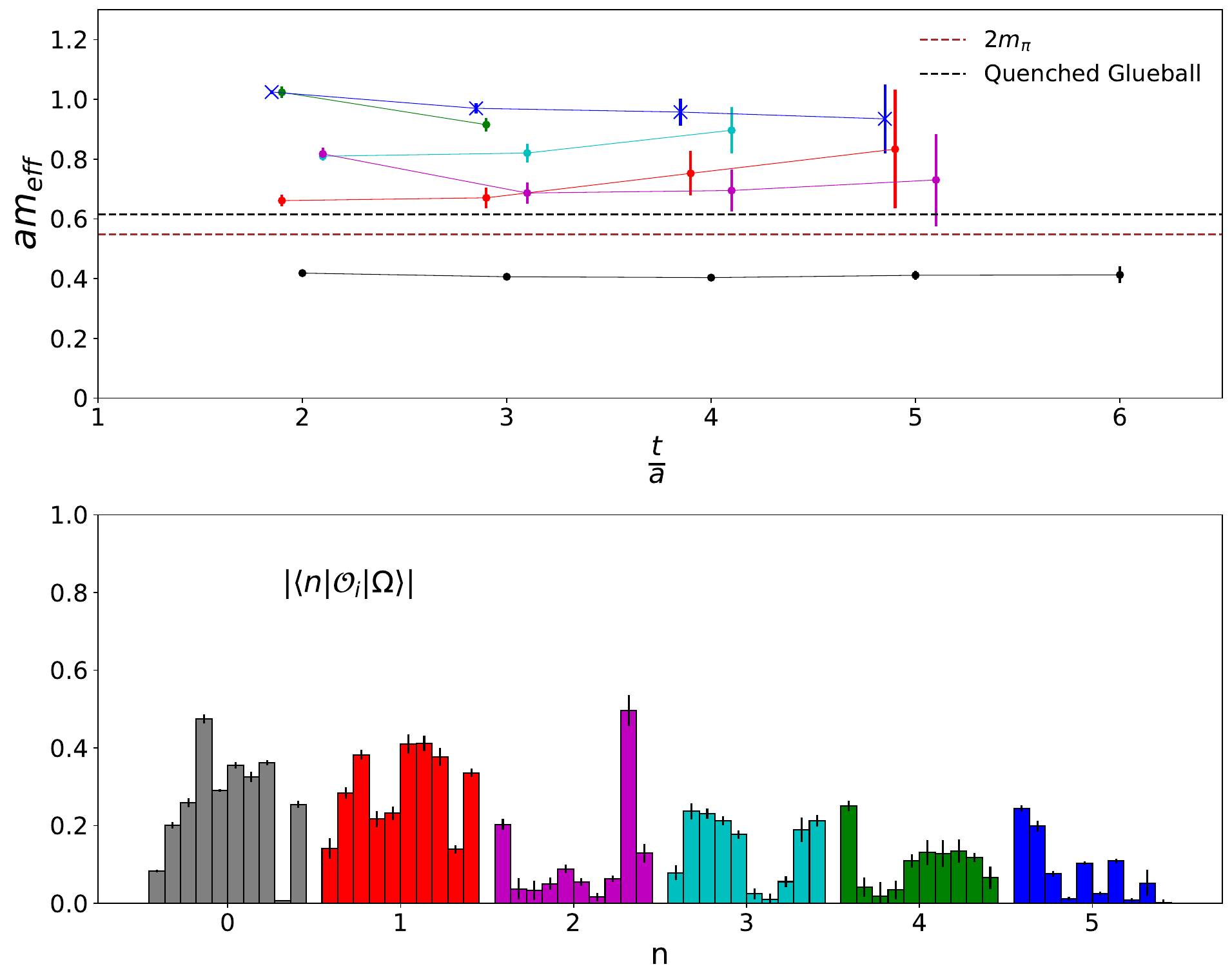}
\caption{\textbf{Upper panel:} Effective masses for $J^{PC} = 0^{++}$ from the GEVP involving all operators from the mixing matrix in ensemble A1h. We only show points with reasonable error bars. The connecting lines are drawn to guide the eye. \textbf{Lower panel:} Normalized overlaps between the states created by each operator and each energy eigenstate. Bars of a same color correspond to the overlaps with a same energy eigenstate. The bars correspond in order to 3 charmonium operators, 5 light meson operators, one 2-pion operator and one gluonic operator.}
\label{fig:Spectrum2piA1heavy}
\end{figure}

The second feature of interest is that no additional low-lying state appears when the gluonic operator is included. This is consistent with the study done in \cite{Brett} and supports the idea that energy eigenstates are a mixture of purely gluonic and mesonic constituent states \cite{Frere, Cheng, Cheng2, Janowski}. The GEVP of the full mixing with the 2-pion operator is shown in the upper panels of  Figs. \ref{fig:Spectrum2piA1} and \ref{fig:Spectrum2piA1heavy}. We found an additional new state, with magenta color points, with respect to the spectrum obtained using 1-particle operators. A reason for this state not appearing in the previous mixings is that 1-particle operators have very little overlap with multi-particle states, so the GEVP can miss such states in the spectrum. The appearance of this additional state indicates how important it is to saturate the spectrum with operators that resemble all possible energy eigenstates in the region of interest. This strategy was used in \cite{Brett} by considering 2-pion, 2-kaon and 2-$\eta$ operators with different values of relative momentum in an $N_f = 2 + 1$ setup where the pions, kaons and $\eta$ are not degenerate. The lower panel of Figs. \ref{fig:Spectrum2piA1} and \ref{fig:Spectrum2piA1heavy} displays the normalized overlaps in absolute value between the states created by the operators we use and the energy eigenstates which are resolved by the GEVP. These are calculated from the GEVP vectors using
\begin{align*}
\bra{\Omega} \mathcal{O}_i \ket{\alpha} &= \left(  U^{\dagger} C(t_0)\right)_i^{\alpha} \sqrt{2m_{\alpha}} e^{\frac{m_{\alpha} t_0}{2}}
\end{align*}
as shown in \cite{Dudek}. Here $U$ has the GEVP vectors as columns, $t_0$ is the reference time of the GEVP, $\left(  U^{\dagger} C(t_0)\right)_i^{\alpha}$ denotes the $i$-th entry of the $\alpha$-th row of $U^{\dagger} C(t_0)$, $\ket{\Omega}$ is the QCD vacuum and $\ket{\alpha}$ is the $\alpha$-th energy eigenstate seen by the GEVP. The overlaps (in absolute value) between a state created by a fixed operator and all the resolved eigenstates are normalized dividing them by their sum. This eliminates the unknown renormalization constant. However, this also means that these overlaps can only be meaningfully compared when they correspond to a same operator but different energy eigenstates. We can see how the states created by light meson operators dominate mostly the low-lying states while the ones created by charmonium operators dominate mostly the higher-lying ones. The state created by the gluonic operator is distributed evenly onto the states below the suspected charmonium one. Only the state created by the 2-pion operators overlaps significantly onto the newly appearing magenta-colored state. 

\section{Conclusions and Outlook}
In this work we performed a first study of the mixing between flavor-singlet 1- and 2-particle operators including light meson, charmonium, purely gluonic and 2-pion ones at two different pion masses. We accounted not only for implicit flavor-mixing via disconnected contributions to the correlation functions, particularly in the charmonium case, but also for explicit mixing by calculating all entries in a mixing correlation matrix. To improve the overlap onto the energy eigenstates of interest we used the improved distillation technique when creating the 1-particle meson operators. In this mixing we observe significant correlations between the light meson, charmonium and gluonic operators. Namely, by including the charm-disconnected contribution we found that the charmonium operators see a state below the expected charmonium ground state. This state is found to be consistent with the states seen by light meson creation operators. If both types of operators are included this state is absorbed by the light meson operators. The excitation which was compatible with the one from connected-only charmonium correlations remained consistent after this change, hinting at it being the true ground charmonium state. The additional inclusion of the gluonic operator left the low-lying spectrum unchanged. Assuming this type of operators couples only to pure glueball states, this further supports results of other calculations in the literature which have found the energy eigenstates are a mixture of pure gluonic and pure mesonic states \cite{Brett, Morningstar}. Finally, the inclusion of a 2-pion operator yielded an additional state in the spectrum which does not couple significantly to the 1-particle operators. This emphasizes the need of multi-particle operators to completely saturate the spectrum in the energy region of interest.

As a next step we will add operators to our basis to better sample the states of interest and make sure we fully saturate the low-lying spectrum. For the 1-particle case we will add operators containing spatial derivatives which can be combined with the improved distillation technique. For the 2-particle case we will use improved distillation as well as include operators with non-zero back-to-back momentum. We are working on applying multi-level sampling to tame the statistical noise in disconnected correlations, which was recently studied in detail for glueball operators in pure-gauge theory \cite{Barca}.

\acknowledgments The authors gratefully acknowledge the Gauss Centre for Supercomputing
e.V. (www.gauss-centre.eu) for funding this project by providing computing time on the GCS
Supercomputer SuperMUC-NG at Leibniz Supercomputing Centre (www.lrz.de) under GCS/LS
project ID pn29se as well as computing time and storage on the GCS Supercomputer JUWELS
at Jülich Supercomputing Centre (JSC) under GCS/NIC project ID HWU35. The authors also
gratefully acknowledge the scientific support and HPC resources provided by the Erlangen National
High Performance Computing Center (NHR@FAU) of the Friedrich-Alexander-Universität
Erlangen-Nürnberg (FAU) under the NHR project k103bf. M. P. was supported by the European
Union’s Horizon 2020 research and innovation programme under grant agreement 824093
(STRONG-2020). R.H. was supported by the programme "Netzwerke 2021", an initiative of the
Ministry of Culture and Science of the State of NorthrhineWestphalia, in the NRW-FAIR network,
funding code NW21-024-A. J. F. acknowledges financial support by the Next Generation Triggers project
(https://nextgentriggers.web.cern.ch). J. A. Urrea-Niño is supported by the German Research Foundation (DFG) research unit FOR5269 "Future methods for studying confined gluons in QCD" and also acknowledges financial support by the Inno4scale project, which received funding from the European High-Performance Computing Joint Undertaking (JU) under Grant Agreement No. 101118139.

\bibliographystyle{JHEP}
\bibliography{refs} 

\end{document}